# Application of stochastic resonance in gravitational-wave interferometer

G.G. Karapetyan

*Cosmic Ray Division, Yerevan Physics Institute, Armenia, 375036*

We investigate novel approach, which improves the sensitivity of gravitational wave (GW) interferometer due to stochastic resonance (SR) phenomenon, performing in additional nonlinear cavity (NC). The NC is installed in the output of interferometer before photodetector, so that optical signal emerging interferometer incidents on the NC and passes through it. Under appropriate circumstances a specific transformation of noisy signal inside the NC takes place, which results in the increase of output signal-to-noise ratio (SNR). As a result optical noisy signal of interferometer becomes less noisy after passing through the NC. The improvement of SNR is especially effective in bistable NC for wideband (several hundreds Hz) detection, when chirp GW signal is detected. Then output SNR is increased up to ~ 0.5 and SNR gain reaches amounts ~ 10. When detection bandwidth is narrowed, the influence of SR mechanism gradually disappears, and SNR gain tends to 1. SNR gain also tends to 1 when the NC is gradually transformed to linear cavity. Proposed enhancement of SNR due to the SR is not dependent of noise type, which is prevalent in interferometer. Particularly proposed approach is the only one, which is capable to increase SNR at given amplitude of displacement noise.

**Introduction**

A number of ground based laser Michelson interferometers are now searching for gravitational waves from different astrophysical objects [1]. Detection of GW is expected to be one of the most exciting scientific results in near future. It will have important applications both for astronomy, where new information about astrophysical objects will be obtained and for fundamental physics, where some aspects of general relativity theory can be tested. Among the most likely detectable sources of the GW are binary systems containing neutron stars and/or black holes [2]. Regardless of the specific configuration, all GW interferometers use the principle of Michelson interferometer to detect the changes $\Delta L = L_1 - L_2$ of the length L in the two perpendicular arm lengths $L_1$ and $L_2$. First generation of GW interferometer having strain sensitivity $\Delta L/L \sim 10^{-21}$ is predicting to detect only large gravitational events. The second generation of detector, such as Advanced LIGO [3] is expected to reach an order higher sensitivity in frequency range several hundreds Hz. These instruments will employ higher power laser, signal recycling technique, more advanced core optics and suspension system. Along with the investigations of these problems, numerous theoretical studies are carried out suggesting novel approaches and principles. These include: the squeezed light [4], quantum demolition technique [5], optomechanical coupling technique [6] Sagnac interferometer [7], white-light cavities [8], stochastic resonance [9], etc. The concept of SR was originally introduced in 1981 [10], then attracting continuously enlarging attention. The SR has been studied in biological systems, information theory, chemical reactions etc. and has been established itself as a large area in noise research. The SR is a specific transformation of noisy signal in nonlinear system, when output signal revealing is improved by the assistance of noise. The response of the system where SR mechanism is performed exhibits resonance-like behavior versus the amplitude of input signal's noise.

In present paper we study application of SR in GW interferometer for improving the sensitivity. To employ the SR phenomenon one needs to have nonlinear system, where noisy signal would be transformed and improved. Earlier in [9] we have considered this problem however, proposed configuration of interferometer was not suitable for experiments. In [9] we suggested to insert optical nonlinear medium in interferometer arm cavities, to change them to nonlinear cavities and to trigger the SR. But such scheme has the drawbacks: (i) because of high power circulating in arm cavities it would be difficult to maintain nonlinear medium in stable state (ii) it would be difficult to provide identical nonlinear parameters for two arm cavities. The scheme suffered also from the necessity to use a special white-light-cavity to relax tight requirements to laser frequency stability.

Here we consider another, simpler configuration where an additional nonlinear cavity (NC) is installed in interferometer output before photodetector. A sketch of proposed configuration is shown in Fig. 1. The function of the NC is to improve SNR of passing through it noisy signal by the SR mechanism. This scheme is easy to realize in all existing ground based interferometers. We present in this paper the results of theoretical study of such interferometer. The main motivation of this work was to attract the attention of GW interferometry experts to the SR, showing on a simple scheme the feasibility of SR application for upgrading the GW interferometers.

Section 1 contains calculations of NC transmittance and SNR of input and output noisy signals. In section 2 it is studied SNR gain for sinusoidal GW signal and in section 3 the SNR gain is studied for chirp GW signal of binary system.

**1. The transmittance of nonlinear FP cavity**

To investigate the possibility of the NC to improve the SNR of noisy signal, one needs to derive the relationship between output ($E_{out}$) and input ($E_{in}$) signal's amplitude. We will use appropriate analytical approach, presented in [9], and based on coupled waves equations [11]. Suppose that the NC is formed by two mirrors, placed at z = 0 and z = D. Region $z_1 < z < z_2$ inside the NC is filled with nonlinear medium with refractive index n = $n_0$ +$n_2$|$E_m$|$^2$, where $n_0$ is background refractive index, $n_2$ is nonlinear coefficient, $E_m$ is the amplitude of electric field of the waves inside nonlinear medium. Output beam of interferometer, having power $P_{in}$ and amplitude of electric field $E_{in}$ incidents on the NC. This beam can be described as a quasi monochromatic plane wave, with electric field given as $E_{in}\exp[i(kz - 2\pi f_{las}t)]$, where k = $2\pi f_{las}/c$ is wave number, $f_{las}$ is laser frequency, c is speed of light in vacuum. Electric field inside the NC is written as

$$\begin{array}{ll} C_1 \exp(ikz) + C_2 \exp(-ikz) & \text{in} \quad 0 < z < z_1 \\ D_1 \exp(ikz) + D_2 \exp(-ikz) & \text{in} \quad z_2 < z < D \\ U(z)\exp(iknz) + V(z)\exp(iknz) & \text{in} \quad z_1 < z < z_2 \end{array} \quad (1)$$

Here $C_1$, $C_2$, $D_1$, $D_2$ are constant complex amplitudes, U(z) and V(z) are slowly varying amplitudes inside nonlinear medium, which obey the coupled waves equations [11]

$$\begin{array}{l} iU'(z) + kn_2 U(z)\left[|U(z)|^2 + 2|V(z)|^2\right] = 0 \\ iV'(z) - kn_2 U(z)\left[|V(z)|^2 + 2|U(z)|^2\right] = 0 \end{array} \quad (2)$$

The solutions of (2) can be presented as

$$\begin{array}{l} U(z) = u \exp[i(\alpha z + \alpha_0)] \\ V(z) = v \exp[i(\beta z + \beta_0)] \end{array} \quad (3)$$

where $\alpha = kn_2(u^2 + 2v^2)$, $\beta = -kn_2(v^2 + 2u^2)$, u and v are real positive constants.

We have 4 boundary conditions for electric and magnetic field at nonlinear medium boundaries, and 4 equations for reflected and transmitted amplitudes at z = 0 and z = D. Thus the system of 8 complex equations is formed, which determines the amplitude of electric field inside and outside the NC.

After appropriate transformations, the following equation connecting output $E_{out}$ and input $E_{in}$ amplitudes is obtained

$$\theta^2 E_{out}^6 + 2\theta\delta t^2 E_{out}^4 + (\delta^2 + t^4)t^4 E_{out}^2 - 4t^8 E_{in}^2 = 0 \quad (4)$$

where $t^2$ = 1 − $r^2$, $\theta$ = $6\pi d(1 + r^2)n_2 f_{las}/c$, d = $z_2$ − $z_1$ is the thickness of nonlinear medium, $\delta$ = $4\pi D(f_{las} - f_{cav})/c$, $f_{cav}$ = $2\pi mc/D$ (m is an integer) is a resonance frequency of empty cavity, closest to laser frequency.

Substituting in (4) a new variable Y as

$$E_{out} = \sqrt{Y - \frac{2t^2\delta}{3\theta}} \tag{5}$$

we come to the cubic equation

$$Y^3 + pY + q = 0 \tag{6}$$

with coefficients

$$p = \frac{t^4(3t^4 - \delta^2)}{3\theta^2}$$
$$q = -\frac{2t^6[\delta(\delta^2 + 9t^2) + 54t^2\theta E_{in}^2]}{27\theta^3} \tag{7}$$

The solutions of equation (6) are given by Cardano's formula

$$Y_1 = a + b$$
$$Y_2 = \frac{a+b}{2} + i\sqrt{3}\frac{a-b}{2} \tag{8}$$
$$Y_3 = \frac{a+b}{2} - i\sqrt{3}\frac{a-b}{2}$$

where

$$a = \sqrt[3]{-\frac{q}{2} + \sqrt{Q}}$$
$$b = \sqrt[3]{-\frac{q}{2} - \sqrt{Q}} \tag{9}$$
$$Q = \left(\frac{p}{3}\right)^3 + \left(\frac{q}{2}\right)^2$$

It can be checked that Q > 0, so $Y_1$ is always real, whereas $Y_2$ and $Y_3$ are real only if p < 0 and $E_1 < E_{in} < E_2$, where

$$E_1 = \frac{\theta}{2t^4}\sqrt{q + \frac{4t^8 E_{in}^2}{\theta^2} - \sqrt{-\frac{4p^3}{27}}}$$
$$E_2 = \frac{\theta}{2t^4}\sqrt{q + \frac{4t^8 E_{in}^2}{\theta^2} + \sqrt{-\frac{4p^3}{27}}} \tag{10}$$

are defined as the lower and upper thresholds of bistable cavity respectively, and $E_0 = (E_1+E_2)/2$. We define $f_0 = f_{cav} - 3^{1/2}t^2c/4\pi D$, then p>=0 is equivalent to $f_{las}$>= $f_0$. In this case $E_{out}$ is described only by solution $Y_1$, being mono value function of $E_{in}$, (see Fig. 2a, b). However if $f_{las} < f_0$ and $E_1 < E_{in} < E_2$, the $E_{out}$ becomes bistable, which means that at given input amplitude $E_{in}$, the $E_{out}$ can get two different values, either in upper or in lower branches (see Fig. 2c). In this case $E_{out}$ is described by both solutions $Y_1$ and $Y_2$, which form together upper and lower branches. This behavior of $E_{out}$ was observed in experiments [12]. Notice that the third solution $Y_3$ describes the middle branch, shown in Fig. 2c by dashed line. This is unstable solution, so $E_{out}$ never is observed in middle branch. The value of $E_{out}$ hops from lower branch to upper one, and vise versa as shown in Fig. 2c, when $E_{in}$ varies with amplitude larger than bistability interval $E_2 - E_1$. If initially $E_{out}$ lies, say on the lower branch, it remains there until $E_{in}$ crosses the upper threshold $E_2$. Then $E_{out}$ hops to upper branch and remains there until $E_{in}$ crosses lower threshold $E_1$. Thus, typical hysteresis phenomenon in $E_{out}$ versus $E_{in}$ relationship is established if $f_0 > f_{las}$. In order to perform effectively the SR mechanism one should have $E_{in}$ varying within bistability interval $E_1 \ldots E_2$. But as it known the GW interferometers work in dark fringe regime, i.e. $E_{in} = 0$ in the absence of GW induced signal. Hence we assume here that in the absence of GW induced signal, $E_{in}$ has a constant mean value $E_{mean}$. This can be provided by a small shift of beam splitter position. We assume that $E_{mean} = E_0 = (E_1+E_2)/2$. Under the action of GW with the strain h(t) the length of one cavity changes on $\Delta L(t) = h(t)L$, while the length of another cavity changes on $-\Delta L(t)$. These changes produce a phase shift $\Delta\Phi(t)$ between two beams reflected from arm cavities. In the result $E_{in}$ becomes

$$E_{in} = E_0 + \Delta E_{in}(t)$$
$$\Delta E_{in}(t) = \Delta E_{GW}(t) = \frac{E_{BS}}{2}\Delta\Phi(t) \quad (11)$$
$$\Delta\Phi(t) = 16F\frac{L}{\lambda}\frac{1}{\sqrt{1+(4\pi\nu\tau)^2}}\frac{\Delta L(t)}{L}$$

where $E_{BS}$ is the amplitude of beam incident on the beam splitter, $F = \pi(r_1 r_2)^{1/2}/(1-r_1 r_2)$ is arm cavity finesse $r_1$, $r_2$ are amplitude reflection coefficients of arm cavity input and rear mirrors, $\lambda = c/f_{las}$ is the wavelength of laser light, $\nu$ is GW frequency, $\tau = 2FL/c$ is arm cavity storage time, $\Delta L(t)$ is the change of arm cavity length L. Taking into account also noise of interferometer, the expression for $\Delta E_{in}(t)$ is written as

$$\Delta E_{in}(t) = \Delta E_{GW}(t) + \xi_{in}(t) \quad (12)$$

where $\xi_{in}(t)$ is noise signal, emerging the interferometer and incident on the NC. This signal contains all components of interferometer noise, including displacement noise, shot noise and photon pressure noise.

To derive SNR of $E_{in}(t)$ one should calculate separately the spectrum $S_{in}(f)$ of signal + noise and the spectrum $N_{in}(f)$ of noise as the following (constant $E_0$ can be omitted in calculations)

$$S_{in}(f) = \int_{t_1}^{t_2} \Delta E_{in}(t)\exp(2\pi i f t)dt$$
$$N_{in}(f) = \int_{t_1}^{t_2} \xi_{in}(t)\exp(2\pi i f t)dt \quad (13)$$

where $t_2 - t_1$ is duration of GW signal.

Then the input SNR, i.e. the SNR of signal emerging interferometer and incident on the NC is determined as

$$SNR_{in} = \left| \frac{\int_{f_1}^{f_2} [S_{in}(f) - N_{in}(f)] df}{\int_{f_1}^{f_2} N_{in}(f) df} \right|^2 \tag{14}$$

where $f_2 - f_1$ is detection bandwidth.

Analogous equation is written for output SNR, i.e. the SNR of signal emerging the NC:

$$SNR_{out} = \left| \frac{\int_{f_1}^{f_2} [S_{out}(f) - N_{out}(f)] df}{\int_{f_1}^{f_2} N_{out}(f) df} \right|^2 \tag{15}$$

where

$$S_{out}(f) = \int_{t_1}^{t_2} E_{out}(t) \exp(2\pi i f t) dt$$
$$N_{out}(f) = \int_{t_1}^{t_2} \xi_{out}(t) \exp(2\pi i f t) dt \tag{16}$$

and $\xi_{out}(t)$ is time variation of noise amplitude at the output of the NC. It is derived from $\xi_{in}(t)$ by the same equations which compute $E_{out}$ versus $E_{in}$.

SNR gain G of noisy signal passing through the NC is therefore

$$G = \frac{SNR_{out}}{SNR_{in}} \tag{17}$$

For further numerical analysis one should specify the parameters of interferometer and the NC. We will use the values close to those of LIGO-1 [13]: L = 4km, $P_{las}$ = 5W, $P_{BS}$ = 150W, F = 100, τ =0.9ms. The value $E_{BS}$ = 7.6kV/m is found from $P_{BS}$ by using the formula $E_{BS} = (960 P_{BS})^{1/2}/d_B$ for uniform circular beam with diameter $d_B$ = 0.05m. Then the dependence between $\Delta E_{GW}(t)$ produced in interferometer output and the change of arm cavity length $\Delta L(t)/L$ under the action of GW with the frequency ν = 150Hz is given numerically as

$$\Delta E_{GW}(t) = 1.3 \cdot 10^{16} \Delta L(t)/L \tag{18}$$

where $\Delta E_{GW}(t)$ is measured in V/m.

To drive the SR mechanism, we need to chouse the parameters of the NC so that $\Delta E_{GW}(t)$ would vary within the bistability interval $E_2 - E_1$.

Key problem in building of the NC is nonlinear medium, which should have rather high value of nonlinear coefficient $n_2$. Recent investigations showed new possibility of creating such a media on the base of Electromagnetic Induced Transparency (EIT) phenomenon [14]. Obtained values of $n_2$ is ~ $10^{-13}$ (m/V)$^2$. We will use this value of $n_2$ as well as the following arbitrary parameters: d = 1cm, r = 0.99, D = 10 cm. NC lengths D is tuned so that $f_0 = f_{las}$ + 119Hz. Then bistable regime in the NC is established with $E_0$ = 63.9054261V/m and $E_2 - E_1 = 2.5*10^{-6}$V/m. Appropriate curve of $E_{out}$ versus $E_{in}$ is shown in Fig. 2c. Note that the instability of beam power

$P_{BS}$, which is necessary to provide the stabile position of $E_{mean}$ close to $E_0$ is ~ $(E_2 - E_1)/E_0$ ~ $4*10^{-8}$, which is easy to realize.

Now we need to determine the function $\xi_{in}(t)$ as well for computation of $S_{in}(f)$ and $N_{in}(f)$ by equations (13). But neither analytical expressions nor experimental data describing time variation of interferometer noise $\xi_{in}(t)$ is known. Instead one should derive $\xi_{in}(t)$ from well known frequency spectrum of interferometer noise by using inverse Fourier transform. We did it for LIGO-1 noise spectrum as the following. First we approximate exact numerical data of LIGO-1 noise spectrum by composing the function F(f)

$$F(f) = 10^{-23}\left(\frac{400000}{f^{2.65}} + 0.015f\right) \quad (18)$$

where f denotes frequency in Hz.

As it is seen in Fig. 3a, the function F(f) well fits exact experimental points of LIGO-1 noise spectrum up to 1000Hz. Then, $N_{in}(f)$ is presented as the product of F(f) and a random function g(f)

$$N_{in}(f) = F(f)g(f) \quad (19)$$

Here g(f) is Gaussian random function with mean 0 and mean square value 0.5. The function $N_{in}(f)$ is shown in Fig. 3b.

Now supposing that $N_{in}(f)$ is noise spectrum of interferometer, we use inverse Fourier transform of $N_{in}(f)$ and obtain time variation $\xi_{in}(t)$ of interferometer noise

$$\xi_{in}(t) = \frac{1}{2\pi}\int_{-\infty}^{\infty} N_{in}(f)\exp(-2\pi ft)df \quad (20)$$

Numerical integration of (20) is done in frequency interval $(-512\ldots512)$Hz by using Fast Fourier Transform (FFT) algorithm with 1024 points. In the result we derived the function $\xi_{in}(t)$, which is close to time variation of LIGO-1 noise (see Fig. 3c).

Slow drift of noise amplitude, seen in Fig. 3c is conditioned by low frequency components in noise spectrum. This drift will deteriorate the performance of the SR because signal amplitude will frequently move outside of bistability interval $E_1\ldots E_2$. To prevent this drift appropriate compensation technique should be used. In such compensation, an error signal is generated, which is proportional to the difference between $E_{mean}$ and $E_0$. This signal is fed back to the actuators to control the amplitude of $E_{mean}$, locking it to $E_0$. Another variant of compensation is to control the resonance frequency $f_{cav}$, controlling by this the bias $\delta$. Then according to (10) $E_1$ and $E_2$ will be controlled, therefore $E_0$ also will be controlled and locked to the $E_{mean}$. The faster is response of compensation loop the higher is cutoff frequency $f_{cut}$ up to which spectral components of interferometer noise $\xi_{in}(t)$ are suppressed. Depending on the quality of compensation scheme, spectral components below $f_{cut}$ can be either completely removed or suppressed. We will assume that compensation scheme provides the suppression of noise spectral components below $f_{cut}$ to the value $F(f_{cut})$. Thus the function $F(f)$ is replaced by another function $F_{comp}(f)$, when compensation scheme is implemented, i.e. instead (19) we have

$$N_{in}(f) = F_{comp}(f)g(f)$$
$$F_{comp}(f) = \begin{cases} F(f) & \text{if } f > f_{cut} \\ F(f_{cut}) & \text{if } f < f_{cut} \end{cases} \quad (21)$$

Computing $\xi_{in}(t)$ according to (20) we obtain time variation of noise amplitude $\xi_{in}(t)$ under the performance of compensation scheme. As it is seen from Fig. 4 the drift of noise $\xi_{in}(t)$ for $f_{cut}$ = 30Hz is about $4*10^{-22}$, which is equivalent to the drift of $E_{mean}$ on about $5.2*10^{-6}$V/m. For $f_{cut}$ = 50Hz these drifts are correspondingly ~ $2*10^{-22}$ and ~ $2.6*10^{-6}$V/m.

## 3. Sinusoidal GW

In this section we investigate SNR gain when GW with sinusoidal waveform is detected. Such GW are emitted by rotating neutron stars or by binary system, long before the coalescence. The change of arm cavity length is presented as

$$\frac{\Delta L(t)}{L} = A\sin(2\pi\nu t) \quad (22)$$

Suppose that $A = 10^{-23}$, $\nu = 150$Hz, and detection bandwidth is $f_2 - f_1 = 450$Hz – 50Hz = 400Hz. Then the noisy signal

$$\Delta E_{in}(t) = 1.3 \cdot 10^{-7} \sin(2\pi\nu t) + \xi_{in}(t) \quad (23)$$

is produced in the output of interferometer according to (11), (12). Signal to noise ratio of this signal, i.e. $SNR_{in}$ is calculated by (14), (13), (12), where $\xi_{in}(t)$ is computed by (20), (21). The calculations are performed as the following.

After generating of 1024 values of random function $g(t)$, by using (21) the noise spectrum $N_{in}(f)$ is calculated, then by using (20) – the function $\xi_{in}(t)$. Substituting $\xi_{in}(t)$ in (23) we calculate time variation of interferometer output signal $\Delta E_{in}(t)$, which is used in (13) to derive the spectrum $S_{in}(f)$ of interferometer output signal. Finally $SNR_{in}$ is computed from (14). But obtained value of $SNR_{in}$ (let define it as $SNR_{in}^{(1)}$) can differ from real experimental value of $SNR_{in}$ because we have used random function $g(f)$. Therefore the procedure of $SNR_{in}$ calculation should be performed again: computation code generates new series of 1024 values of function $g(f)$ and new $SNR_{in}^{(2)}$ is derived. Repeating this procedure many times we obtain different values of $SNR_{in}^{(i)}$, i = 1,2…K. Then the robust value of $SNR_{in}$ is obtained by averaging these values:

$$SNR_{in} = \frac{1}{K}\sum_{i=1}^{K} SNR_{in}^{(i)} \qquad (24)$$

Number of trials K should be choused so that adding of new value $SNR_{in}^{(K+1)}$ does not significantly change $SNR_{in}$ in (24). As it is seen from Fig. 5, it is sufficient to take K ~ 30. Then the value of $SNR_{in}$ proves to be ~0.21. The $SNR_{out}$ is calculated analogously by using the equations (15) and (16). However it is necessary to use larger number K, so we took K = 1000.

First let us investigate the dependence of G from the bistability interval $E_2 - E_1$. As it was mentioned earlier this interval is determined by the bias between laser frequency $f_{las}$ and $f_0$ according to (10), (7). Hence the change of the interval $E_2 - E_1$ can be provided by the change of $f_{cav}$, which can be done by tuning of the NC length D. In Fig. 6 the SNR gain G versus the length of bistability interval is plotted. It is seen significant gain ~5 at the value $E_2 - E_1 \sim 2.5*10^{-6}$V/m. When $E_2 - E_1$ surpasses $2.5*10^{-6}$V/m, the gain sharply drops to ~1. It also decreases to 1 when $E_2 - E_1$ tends to 0 and NC becomes monostable. Such resonance behavior of SNR gain versus bistability interval is just caused by the SR mechanism, which performs here as the following.

When $E_2 - E_1 \gg \Delta E_{in}(t)$ (in our case $\Delta E_{in}(t) \sim 2.5*10^{-6}$V/m ), the $E_{out}(t)$ is localized on one of the branches, so the relationship between $E_{out}$ and $E_{in}$ is approximately linear, and G ~ 1. By decreasing $E_2 - E_1$ this linear relationship preserves until $E_2 - E_1 \sim \Delta E_{in}(t)$. But as soon as $\Delta E_{in}(t)$ begins to cross the upper $E_2$ and lower $E_1$ thresholds and $E_{out}$ begins to hop from one branch to another one, the ratio $E_{out}/E_{in}$ sharply increases. Although crossing of the thresholds and the hops of $E_{out}$ takes place due to the net input noisy signal $\Delta E_{in}(t)$, nevertheless composition of GW induced signal in $E_{out}$ increases resulting to the increase of $SNR_{out}$ and G. Further decrease of bistability interval leads to the situation when $E_2 - E_1 \ll \Delta E_{in}(t)$. Now the influence of hops in changing the ratio $E_{out}/E_{in}$ weakens, resulting in reduction of SNR gain. Notice that at $E_2 - E_1 = 0$ (when the NC becomes monostable), the small SNR gain is still observed, i.e. the SR mechanism performs in monostable nonlinear cavity as well, but with the less efficiency.

Although noise signal $\xi_{in}(t)$ in our study has definite amplitude, determined by above used noise spectrum of LIGO-1, nevertheless it would be useful to investigate the behavior of $SNR_{out}$ when noise amplitude is changed. For this let us replace $\xi_{in}(t)$ in (12), (13) by the function $B\xi_{in}(t)$, where B is dimensionless constant coefficient. As it known and as it follows from (14) the dependence of $SNR_{in}$ versus noise amplitude B is $\sim 1/B^2$, i.e. $SNR_{in}$ monotonically decreases when noise amplitude increases. However the $SNR_{out}$ essentially differs from $SNR_{in}$ due to SR mechanism. To derive the behavior of $SNR_{out}$ let us specify again the bistability interval $E_2 - E_1$ = $2.5*10^{-6}$V/m, corresponding to maximum G at B = 1. Then by changing B we obtain the dependence of $SNR_{out}$ versus the amplitude of input noise. The computations are performed with the number of trials K = 1000. In the result we obtained the graphs presented in Fig. 7. We can detect the resonance like behavior of SNR$_{out}$ versus input noise amplitude, which is the signature of SR mechanism performing here. SNR$_{out}$ ~ SNR$_{in}$ when noise amplitude is small. However by increasing of the B when ΔE$_{in}$(t) begins to cross the thresholds and E$_{out}$ begins to hop from one branch to another one, the SNR$_{out}$ sharply increases, surpassing SNR$_{in}$ at the maximum point by ~ 5 times. One can see also that the SNR gain is reduced when detection bandwidth decreases.

## 4. Chirp GW

In this section we study the SNR gain for the case of GW with chirped waveform. Such GW is generated by compact binary system in the final stage of inspiral, when the stars merge into coalesced state. GW strain can be presented as the following [14]

$$h(t) = \frac{A}{\sqrt[4]{t_{coal} - t}} \cos\left\{\left[\frac{c^3}{5\gamma M}(t_{coal} - t)\right]^{5/8}\right\} \quad (25)$$

where t$_{coal}$ is coalescence time, γ = 6.67*10$^{-11}$n*m$^2$/kg$^2$ is gravitational constant, M=(M$_1$M$_2$)$^{3/5}$/(M$_1$+M$_2$)$^{1/5}$, M$_1$ and M$_2$ are the masses of the two stars.

Suppose that both of stars has solar mass, M$_1$ = M$_2$ = M$_{sun}$ = 1.99*10$^{30}$kg, and the amplitude A = 7*10$^{-24}$. In Fig. 8 there are presented the waveform of h(t) at last ~ 1 sec before coalescence, and its spectrum. The detection of this GW signal by LIGO-1 with detection bandwidth f$_2$ – f$_1$ = 450Hz – 50Hz = 400Hz is non effective, because according to (12), (13) and (14) SNR$_{in}$ = 0.05. However by passing this noisy signal through the NC with the same parameters as in previous section, the output SNR can be increased due to the SR mechanism. Again let us investigate first the dependence of SNR gain from bistability interval. As it is seen in Fig. 9 the SNR gain is G ~ 10 for cutoff frequency f$_c$ = 50Hz and detection bandwidth f$_2$ – f$_1$ = 400Hz. By decreasing detection bandwidth, the SNR gain decreases tending to 1 analogously to previous case of sinusoidal GW, so these graphs are not shown here. Instead we present the graphs corresponding to different cutoff frequencies to see how SNR gain deteriorates when cutoff frequency of compensation scheme decreases. It is seen that bistability interval where SNR gain is observed increases, and maximal SNR gain reduces when cutoff frequency decreases. However, even at f$_c$ = 10Hz the SR mechanism successfully performs here, providing SNR gain ~ 6.5. Calculations of SNR$_{out}$ were made analogously to those of previous section, with number of trials K = 1000. Finally we investigated SNR$_{out}$ versus the amplitude of input noise, by replacing analogously of previous case, input noise amplitude ξ$_{in}$(t) by Bξ$_{in}$(t). In the result analogous graphs representing the dependence of SNR$_{out}$ versus B is derived (Fig. 10). It is seen resonance like behavior of SNR$_{out}$, which provides the increase of SNR$_{out}$ to value ~ 0.5 (SNR gain is G ~ 10). Thus SR mechanism performs more effective when detecting chirp GW signal comparing with sinusoidal one.

In conclusion we investigated application of SR phenomenon in GW interferometer for increasing the sensitivity. It is done by passing output signal through a nonlinear cavity installed in the output of interferometer. When appropriately tuned, the NC becomes capable to increase the SNR of passing through it noisy signal due to SR phenomenon. We performed detail computations of SNR gain for two types of GW waveform – sinusoidal and chirp, taking as an example the parameters of interferometer close to those of LIGO-1. It is obtained that the influence of SR is most effective for wideband (several hundreds Hz) detection when bistable regime of the NC is employed. Then $SNR_{out}$ is ~ 0.5 and G ~ 10 for chirp GW signal. Proposed approach can be used for upgrading any GW interferometer by minor changes in their configuration.

**Acknowledgements**

I thank Prof. A. Chilingarian for fruitful discussions and PhD student A Hakobyan for the help in figures preparation.
This work was supported by the Armenian National Science and Education Fund.

**Figure captions**

Fig. 1. Schematic of proposed interferometer's configuration. PRM – power recycling mirror, BS – beam splitter, NC – nonlinear cavity, PD – photodetector, $P_{BS}$ and $E_{BS}$ – power and amplitude of electric field incident on beam splitter, $E_{in}$ – amplitude of the beam incident on the NC, $E_{out}$ – amplitude of the beam emerging the NC.

Fig. 2. Output amplitude of light passing through nonlinear cavity versus the input amplitude. D = 10cm, d = 1cm, $n_2 = 10^{-13}(m/V)^2$, r = 0.99, $E_0$ = 63.9054261 V/m.

Fig. 3. a) Discrete values of LIGO-1 noise spectrum (circles) and function F(f), approximating these values (solid line); b) Composed noise spectrum $N_{in}(f)$, which is close to LIGO-1 noise spectrum; c) Time variation of noise amplitude $\xi_{in}(t)$, which is close to time variation of LIGO-1 noise.

Fig. 4. Time variation of noise amplitude $\xi_{in}(t)$ at different cutoff frequencies of compensation scheme.

Fig. 5. Computed values of $SNR_{in}$ versus the number of trials K for sinusoidal GW with frequency 150Hz and strain h = $2*10^{-23}$. Parameters of interferometer are close to those of LIGO-1, D = 10cm, d = 1cm, $n_2 = 10^{-13}(m/V)^2$, r = 0.99, detection bandwidth is $f_2$ = 450Hz, $f_1$ = 50Hz, cutoff frequency is $f_c$ = 50Hz.

Fig. 6. SNR gain versus bistability interval at different detection bandwidths, when detecting the sinusoidal GW with frequency f = 150Hz and strain h = $2*10^{-23}$. Parameters of interferometer are close to those of LIGO-1, D = 10cm, d = 1cm, $n_2 = 10^{-13}(m/V)^2$, r = 0.99, $E_0$ = 63.9054261 V/m, $E_2 - E_1 = 2.5*10^{-6}$V/m, cutoff frequency is $f_c$ = 50Hz.

Fig. 7. Output SNR versus the amplitude of input noise (circles + line) at different detection bandwidths when sinusoidal GW with frequency f = 150Hz and strain h = $2*10^{-23}$ is detected. Parameters of interferometer are close to those of LIGO-1, D = 10cm, d = 1cm, $n_2 = 10^{-13}(m/V)^2$, r = 0.99, $E_0$ = 63.9054261 V/m, $E_2 - E_1 = 2.5*10^{-6}$V/m, $f_c$ = 50Hz. Value B = 1 corresponds to noise amplitude shown in Fig. 4a, which is close to LIGO-1 noise. The dashed line represents $SNR_{in}$.

Fig. 8. Waveform and the spectrum of chirp GW emitted by coalesced binary system with the masses $M_1 = M_2 = M_{sun}$, at last ~ 1sec before coalescence.

Fig. 9. SNR gain versus bistability interval at different cutoff frequencies, when detecting the chirp GW with $A = 7*10^{-24}$. Parameters of interferometer are close to those of LIGO-1, $D = 10$cm, $d = 1$cm, $n_2 = 10^{-13}(m/V)^2$, $r = 0.99$, $E_0 = 63.9054261$ V/m, $E_2 - E_1 = 2.5*10^{-6}$V/m, detection bandwidth is $f_2 = 450$Hz, $f_1 = 50$Hz.

Fig. 10. Output SNR versus the amplitude of input noise (circles + line) at different detection bandwidths when chirp GW is detected. Parameters of interferometer are close to those of LIGO-1, $D = 10$cm, $d = 1$cm, $n_2 = 10^{-13}(m/V)^2$, $r = 0.99$, $E_0 = 63.9054261$ V/m, $E_2 - E_1 = 2.5*10^{-6}$V/m, $f_c = 50$Hz. Value $B = 1$ corresponds to noise amplitude shown in Fig. 4a. The dashed line represents $SNR_{in}$.

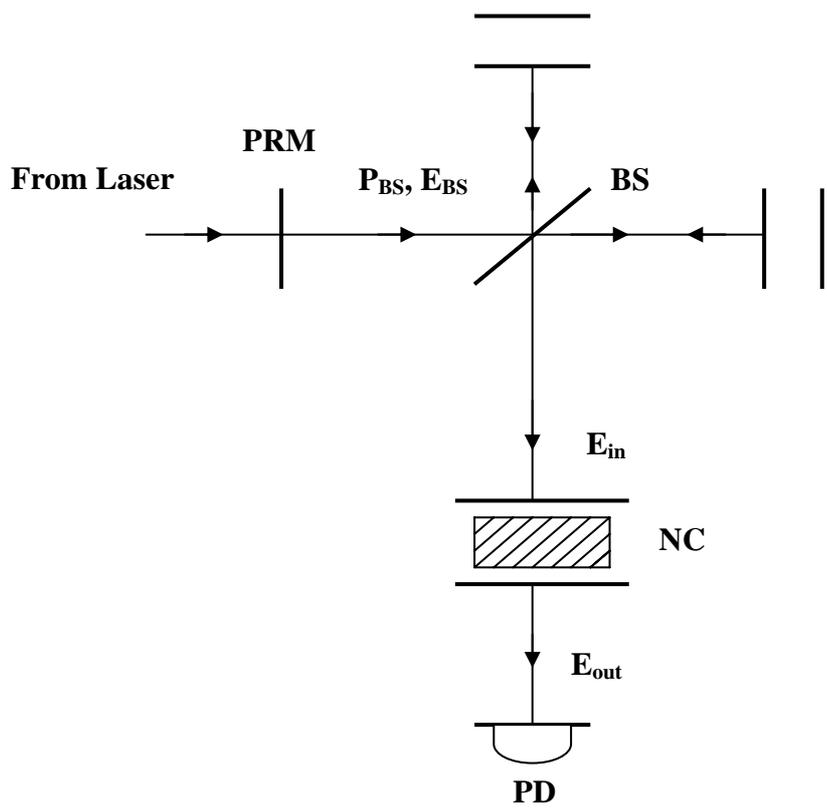

Fig. 1

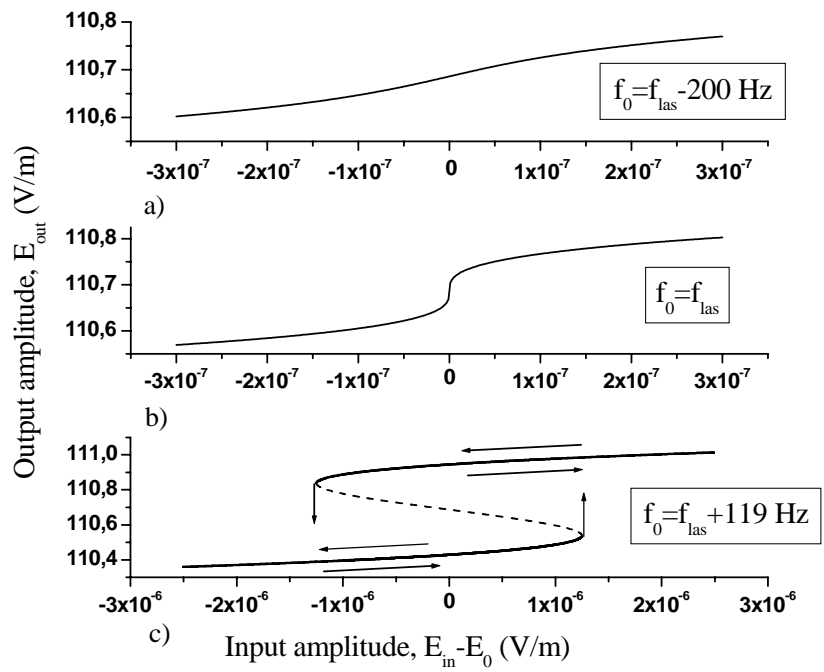

Fig. 2

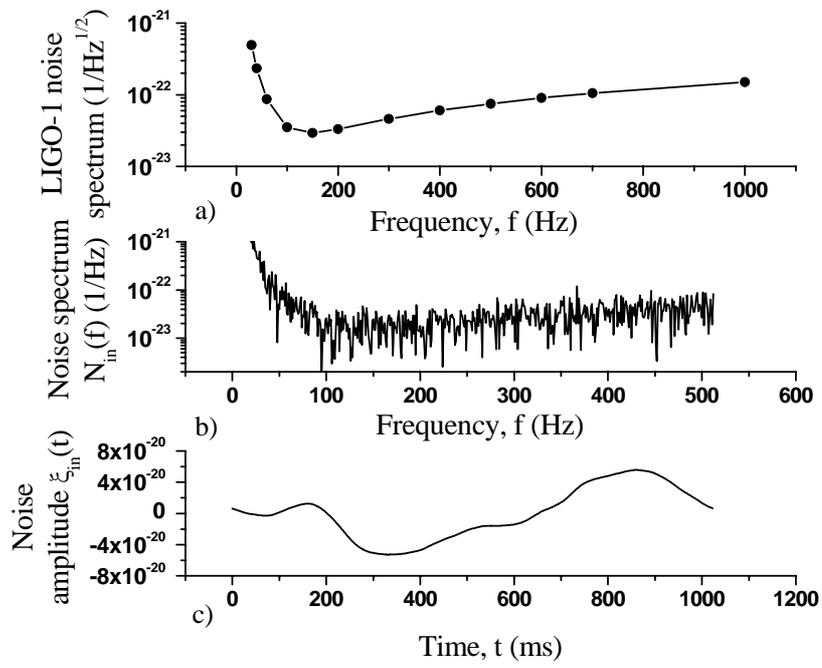

Fig. 3

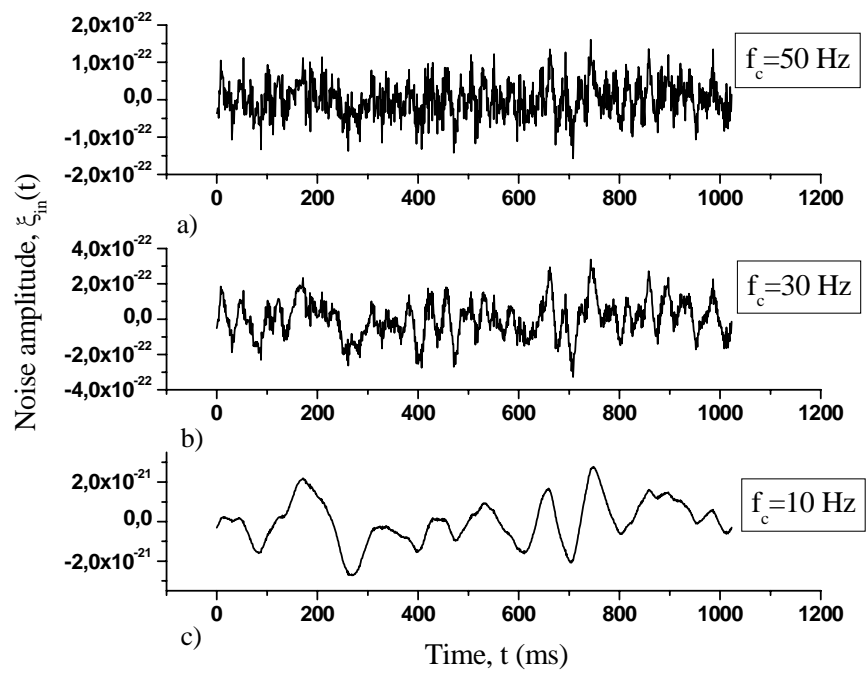

Fig. 4

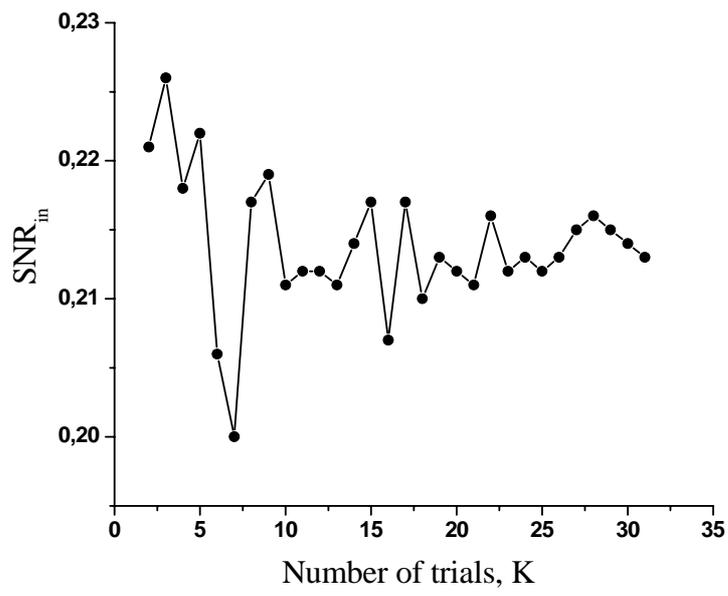

Fig. 5

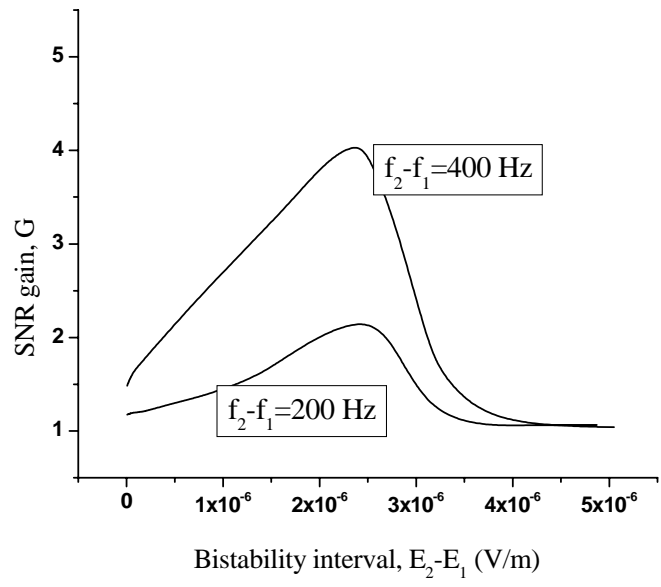

Fig. 6

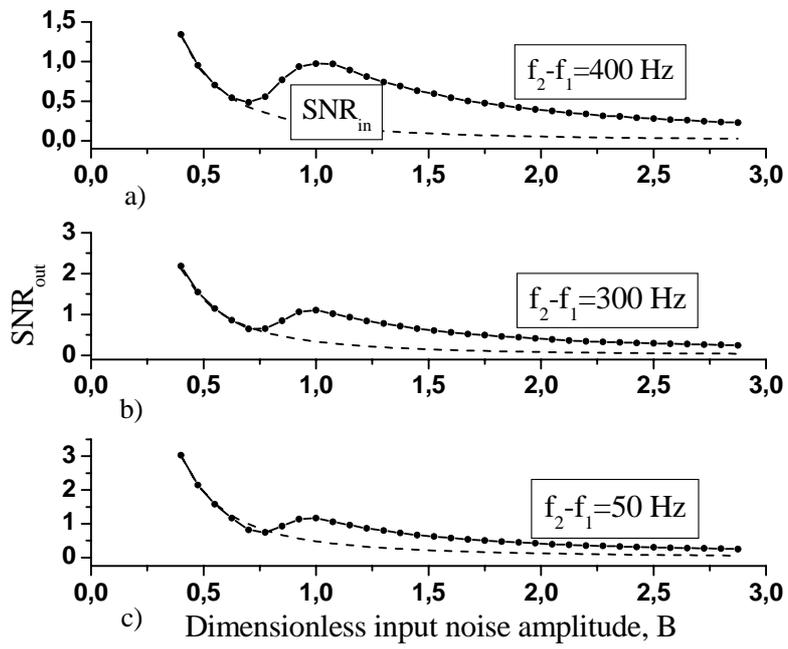

Fig. 7

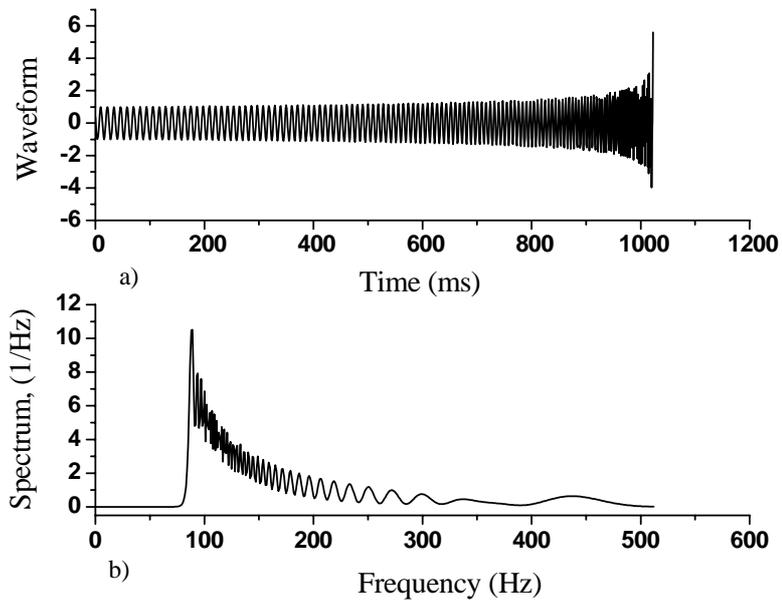

Fig. 8

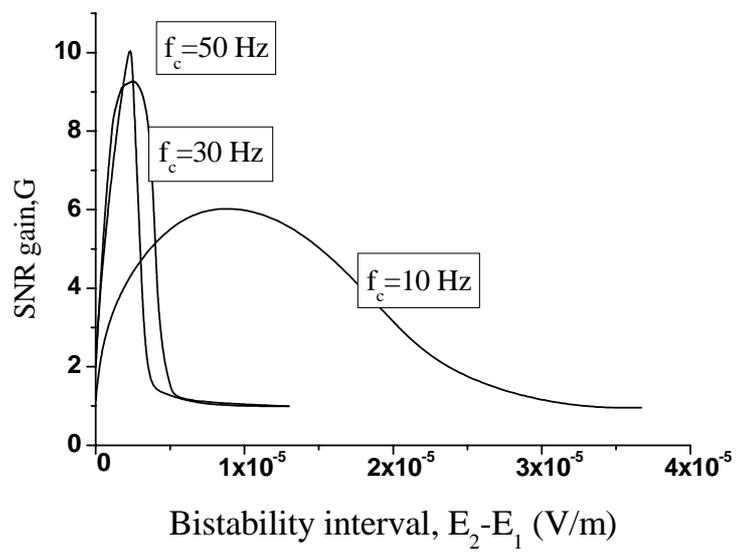

Fig. 9

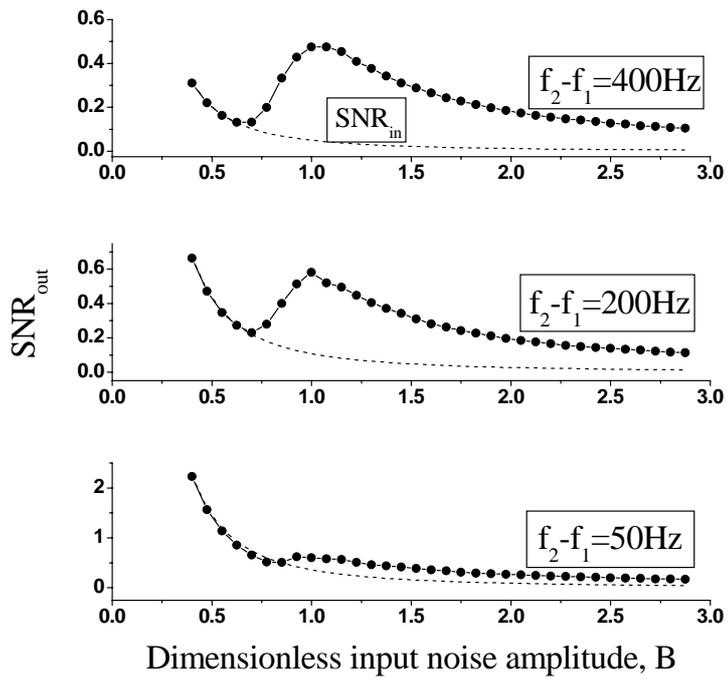

Fig. 10